\RequirePackage{ifpdf}
\ifpdf
	\documentclass[twocolumn, prl, showkeys, letterpaper, pdftex]{revtex4}
\else
	\documentclass[twocolumn, prl, showkeys, letterpaper, dvips]{revtex4}
\fi

\usepackage{amsmath}
\usepackage[nodvipsnames]{color}

\ifpdf
  \usepackage[pdftex]{graphicx}
  \usepackage{subfigure}
  \graphicspath{{pdf-figures/}}
  \usepackage[%
    pdftex,%
    pdftitle = {{Interplay between diffusive and displacive phase transformations: TTT diagrams and microstructures}},%
	pdfsubject = {{Materials which can undergo extremely fast displacive transformations as well as very slow diffusive transformations are studied using a Ginzburg-Landau framework to understand the physics behind microstructure formation and TTT diagrams.}},%
	pdfauthor = {{Mathieu Bouville and Rajeev Ahluwalia}},%
	pdfkeywords = {{spinodal decomposition, phase transformation, TTT diagram, phase-field, steel, martensite, pearlite, austenite, bainite, elastic compatibility}},%
  ]{hyperref}
\else
  \usepackage[dvips]{graphicx}
  \usepackage{subfigure}
  \usepackage[
	dvips,
    pdftitle = {{Interplay between diffusive and displacive phase transformations: TTT diagrams and microstructures}},%
	pdfsubject = {{Materials which can undergo extremely fast displacive transformations as well as very slow diffusive transformations are studied using a Ginzburg-Landau framework to understand the physics behind microstructure formation and TTT diagrams.}},%
	pdfauthor = {{Mathieu Bouville and Rajeev Ahluwalia}},%
	pdfkeywords = {{spinodal decomposition, phase transformation, TTT diagram, phase-field, steel, martensite, pearlite, austenite, bainite, elastic compatibility}},%
	letterpaper,%
]{hyperref}
\fi

\newcommand{\p}{\ensuremath{_\text{P}}}
\newcommand{\m}{\ensuremath{_\text{M}}}

\begin{document}

\title{Interplay between diffusive and displacive phase transformations:\\TTT diagrams and microstructures}

\author{Mathieu Bouville}
	\email{m-bouville@imre.a-star.edu.sg}
	\affiliation{Institute of Materials Research and Engineering, Singapore 117602}

\author{Rajeev Ahluwalia}%
	\email{a-rajeev@imre.a-star.edu.sg}
	\affiliation{Institute of Materials Research and Engineering, Singapore 117602}

\date{\today}

\begin{abstract}
Materials which can undergo extremely fast displacive transformations as well as very slow diffusive transformations are studied using a Ginzburg--Landau framework to understand the physics behind microstructure formation and time--temperature--transformation (TTT) diagrams. This simple model captures the essential features of alloys such as steels and predicts the formation of mixed microstructures by an interplay between diffusive and displacive mechanisms. The intrinsic volume changes associated with the transformations stabilize mixed microstructures such as martensite--\linebreak[3]retained austenite (responsible for the existence of a martensite finish temperature) and martensite--pearlite.
\end{abstract}
\maketitle


Based on kinetic considerations, solid-to-solid phase transformations can be broadly categorized as diffusive and displacive~\cite{Christian-book-02}.
Diffusive phase transformations ---such as spinodal decomposition--- involve long-range motion of atoms and are therefore slow. In displacive phase transformations ---e.g.\ martensitic transformations~\cite{Bhattacharya-book}--- the crystal structure changes through a unit cell distortion, without any long-range motion of the atoms.

These two types of phase transformations may interact or compete, e.g.\ in steel~\cite{Honeycombe-book}, Ti--Al--Nb~\cite{Ren-acta_mater-01}, Cu--Al--Ag~\cite{Adorno-J_alloys_Comp-01}, Pu--Ga~\cite{Hecker}, and NiTi shape memory alloys~\cite{Gall-acta_mater-02}. Eutectoid steels decompose into pearlite, i.e.\ ferrite plus cementite (iron carbide, Fe$_3$C), by a diffusive process. However, at low temperature carbon diffusion is very slow and, although it is not the ground state, martensite can form on fast cooling. At intermediate temperature a different microstructure, bainite, may form in steel through a mixed diffusive--displacive mechanism~\cite{Bhadeshia-bainite_book}. Information on these phase transformations and their kinetics can be conveniently represented on time--transformation--temperature (TTT) diagrams. A theoretical understanding of the interplay between diffusive and displacive modes is paramount to comprehend TTT diagrams and microstructures, which are important technologically in the design of heat treatments.

\Citet{Olson-acta_met-89} theoretically  studied coupled diffusive and displacive transformations to calculate a TTT diagram for bainite. However, this one-dimensional model cannot describe the microstructural complexity and the long-range elastic fields generated during such transformations. The Ginzburg--Landau method (also known as the phase-field method) on the other hand can naturally describe the appropriate habit planes without assuming any given microstructure.
It has been extensively used to study microstructural evolutions in diffusive~\cite{Cahn-acta_met-61, Halperin-Rev_Mod_Phys-77, Bray-94, Onuki-PRL-01} and martensitic~\cite{Falk80, Wang97, Onuki-99, Vedantam-05, Ahluwalia-acta_mater-06} transformations, but (apart from one work on a TTT diagram for ferrite and martensite in iron~\cite{Rao-prl-05}) these phase transformations have typically been studied separately.

Using the Ginzburg--Landau theory, we study a model system which can undergo both a phase separation and a square-\linebreak[3]to-\linebreak[3]rectangle martensitic transformation. We label ``pearlite'' {\itshape any} phase-separated region, even if the microstructure does not correspond to real pearlite {\it per se}. 
Although this letter focuses on TTT diagrams, other thermal treatments are possible using this model, e.g.\ continuous cooling transformation (CCT). 
Likewise martensitic transformations other than the square-\linebreak[3]to-\linebreak[3]rectangle studied here can be modeled.


The free energy of the system is expressed as
\begin{equation*}
	G=\int \left(g_\text{el} + g_\text{ch} + g_\text{cpl} + \frac{k_c}{2} \|\nabla c\|^2 + \frac{k_{e2}}{2} \|\nabla e_2\|^2 \right) \mathrm{d}\mathbf{r}.
\end{equation*}
\noindent Here $g_\text{el}$ is the usual non-linear elastic free energy density for a square-\linebreak[3]to-\linebreak[3]rectangle martensitic transition \cite{Falk80, Onuki-99}: 
\begin{eqnarray}
	&g_\text{el} = &\dfrac{A_2}{2}\dfrac{T\!-T\m}{T\m}e_2^2 - \dfrac{A_4}{4}e_2^4 + \dfrac{A_6}{6}e_2^6 +\nonumber\\
	& & \dfrac{B_1}{2} \left[ e_1 - \left( x_{1c}\,c + x_{12}\,e_2^2  \right)\right]^2 + \dfrac{B_3}{2} e_3^2,
\label{eq-g_el}
\end{eqnarray}
\noindent where $e_1 = (\varepsilon_{xx}+\varepsilon_{yy})/\sqrt{2}$ is the hydrostatic strain, $e_2 = (\varepsilon_{xx}-\varepsilon_{yy})/\sqrt{2}$ is the deviatoric strain, and $e_3 = \varepsilon_{xy}$ is the shear strain. The $\{\varepsilon_{ij}\}$ are the linearized strain tensor components. $T$ is the temperature and $T\m$ and $T\p$ are constants pertaining to the austenite--martensite and austenite--pearlite phase transformations respectively. $c$ is the composition ($c=0$ corresponds to the composition of austenite and the composition of pearlite is $c=\pm c_0$).

The chemical free energy is given by~\cite{Cahn-acta_met-61}
\begin{equation}
	g_\text{ch} = \dfrac{C_2}{2}\dfrac{T\!-T\p}{T\p}c^2 + \dfrac{C_4}{4}\, c^4
\label{eq-g_chem}
\end{equation}
\noindent and a coupling between elastic distortions and composition is introduced as $g_\text{cpl} = x_{2c}\, c^2 e_2^2$.
Notice that in the stress-free conditions, at equilibrium $e_1=x_{1c}\,c+x_{12}\,e_2^2$: during the transformations, volume changes may be introduced, either due to a lattice mismatch between the phase-separating components ($x_{1c}$) or due to the transformation strains ($x_{12}$).

\begin{figure}
\centering
\setlength{\unitlength}{1cm}
\begin{picture}(8.5,3.15)(.2,0)
\subfigure
{
    \label{energy(a)}
    \includegraphics[height=3.15cm]{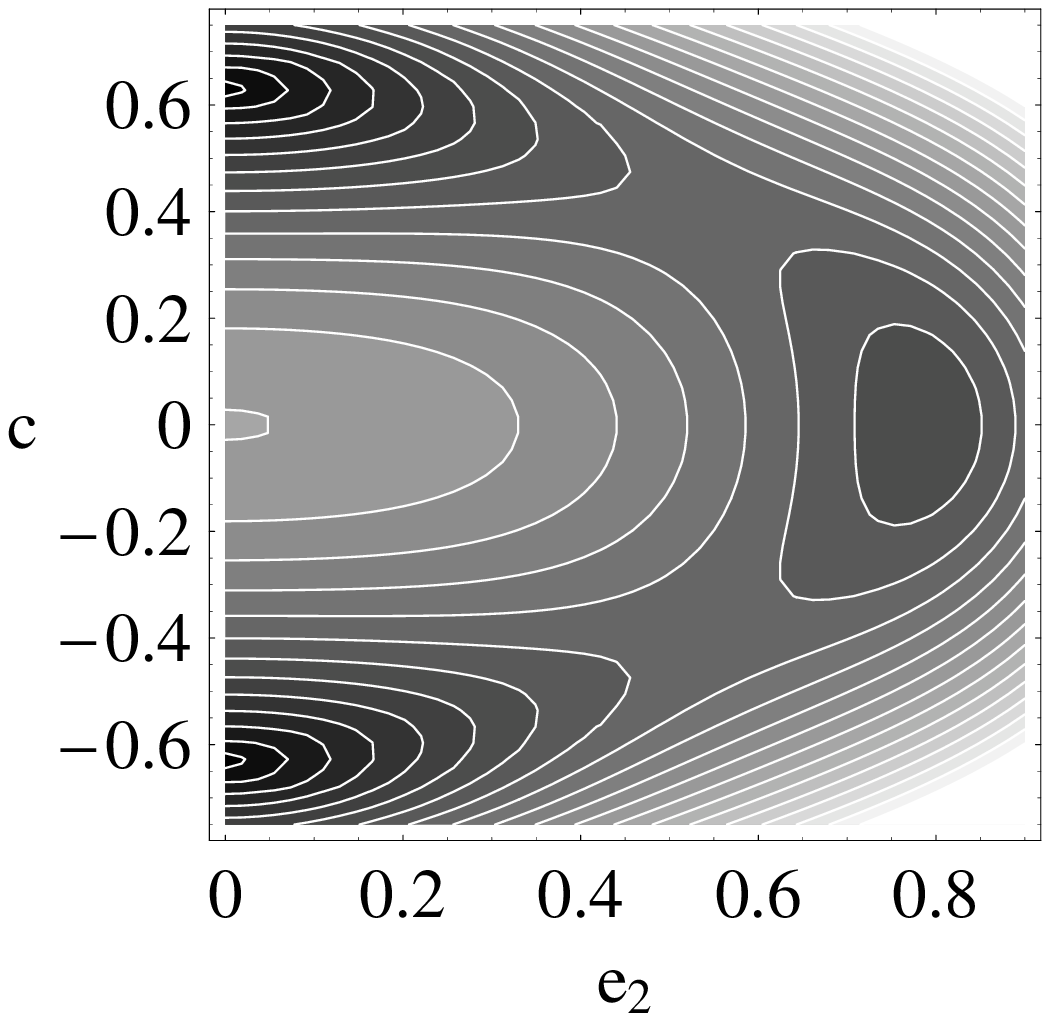}
	\put(-0.65, 2.75){\bf(a)}
}\subfigure{
    \label{energy(b)}
    \includegraphics[height=3.15cm]{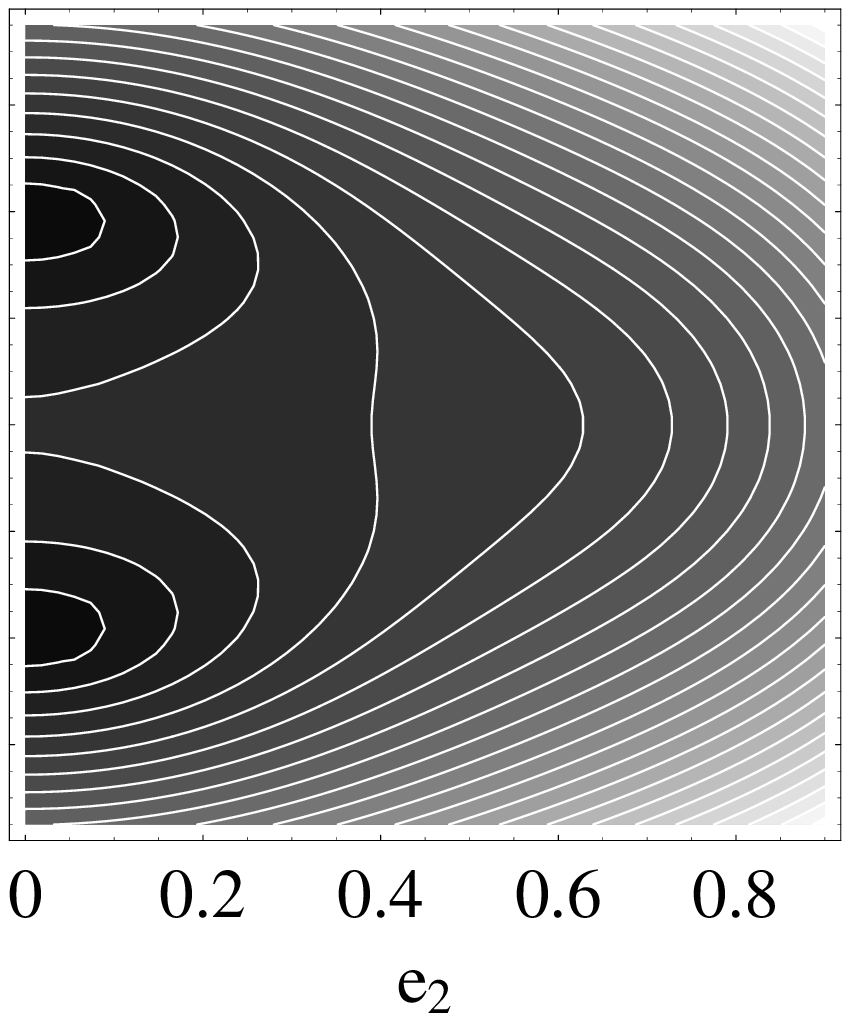}
	\put(-0.65, 2.75){\bf(b)}
}\subfigure{
    \label{energy(c)}
    \includegraphics[height=3.15cm]{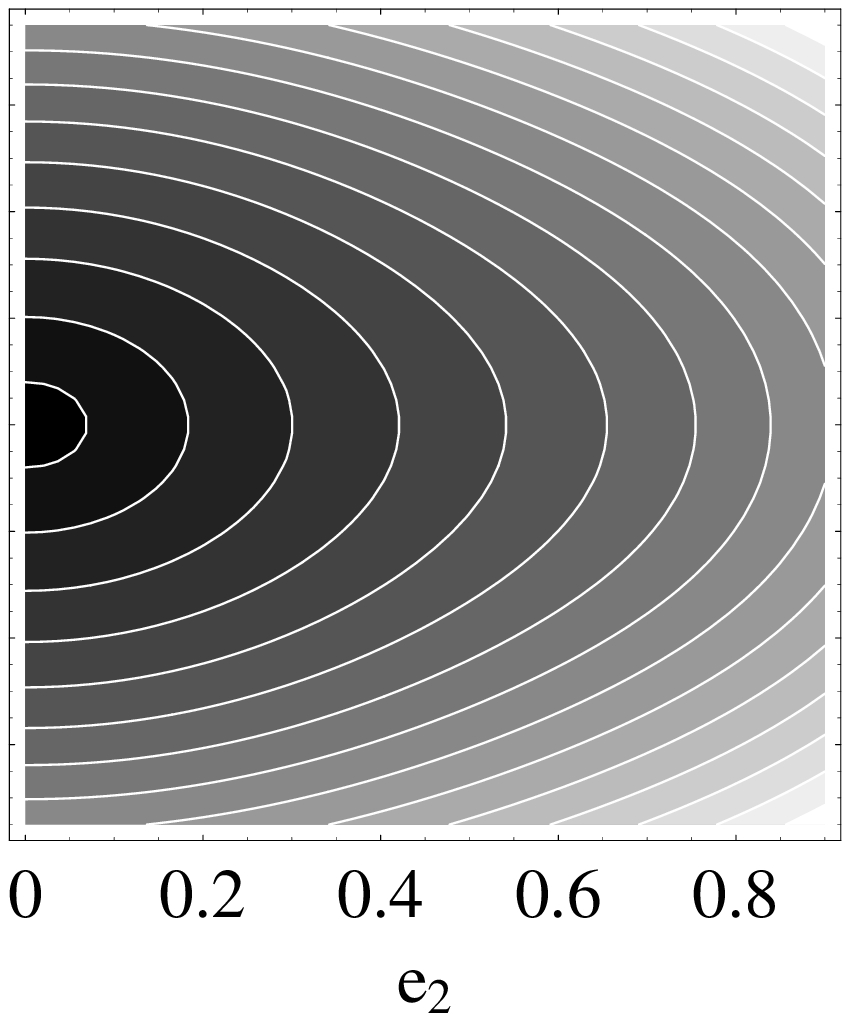}
	\put(-0.65, 2.75){\bf(c)}
}
\end{picture}
\caption{\label{energy}The bulk energy as a function of $e_2$ and $c$ at (a) $T = 0.2$, (b) $T = 0.7$, and (c) $T = 1.5$. Darker areas correspond to lower energies.}
\end{figure}

Since we are interested in a qualitative understanding of the physical mechanisms, we choose simple values for the parameters: $A_2=2$, $A_4=4$, $A_6=9.6$, $B_1=1$, $B_3=1$, $C_2=6$, $C_4=12$, $x_{2c}=5$, $T\m=0.5$, $T\p = 1$, $k_c=2$, and $k_{e2}=0.1$.
The coupling constants $x_{12}$ and $x_{1c}$ are varied in different cases to understand the effect of volume changes. 
The homogeneous part of the free energy is depicted in Fig.~\ref{energy} as a function of $e_2$ and $c$ at different temperatures. 
The different phases can be identified as follows: austenite corresponds to $({c=0},\,{e_2=0})$, martensite to $({c=0},\,{e_2\ne0})$, and pearlite to $({c\ne0},\,{e_2=0})$.
\noindent Above $T\p$ only austenite is stable, Fig.~\ref{energy(c)}. Between $T\m$ and $T\p$ austenite and martensite are unstable and pearlite is the ground state, Fig.~\ref{energy(b)}. Below $T\m$, pearlite is the ground state and martensite is metastable, Fig.~\ref{energy(a)}. 

The kinetics of the transformations are described by equations of motion for the displacement fields and the composition. The evolution of the displacements is described by~\cite{landau-lifschitz, Ahluwalia-acta_mater-06}
\begin{subequations}
\begin{align}
	\rho \,\frac{\partial^2\, u_i(\mathbf{r}, t)}{\partial\, t^2} &= 
\sum_j\frac{\partial\, \sigma_{ij}(\mathbf{r}, t)}{\partial\, r_j} +
\eta \,\mathbf{\nabla}^2 v_i(\mathbf{r}, t),
\label{du_dt}\\
	\sigma_{ij}(\mathbf{r}, t) &= \frac{\delta\, G}{\delta\, \varepsilon_{ij}(\mathbf{r}, t)},
\end{align}
\end{subequations}
\noindent where $\delta$ is the functional derivative, $\rho$ is a density, and $\mathbf{v}$ is the time derivative of the displacements, $\mathbf{u}$. The second term on the right-hand side in Eq.~(\ref{du_dt}) is a viscous damping term. The strains, $\{\varepsilon_{ij}\}$, are then obtained as derivatives of the displacements.
The evolution of the composition is described through the Cahn--Hillard equation~\cite{Cahn-acta_met-61}
\begin{equation}
	\frac{\partial\, c(\mathbf{r}, t)}{\partial\, t} = M\, \mathbf{\nabla}^2 \frac{\delta\, G}{\delta\, c(\mathbf{r}, t)},
\end{equation}
\noindent where $M$ is the temperature-dependent mobility,
\begin{equation}
	M = M_0 \,\exp(-Q/T).
\end{equation}
\noindent Note that this temperature dependence of this mobility is 
necessary in order to obtain realistic TTT diagrams.

All simulations are two-dimensional with periodic boundaries and on a $128 \times 128$ lattice ($\delta x=1$ and $\delta t=0.2$). We use $\eta = 0.01$, $\rho =1$, $M_0 = 2$, and $Q = 5$.
The initial system, made of 100\% austenite, includes random fluctuations around $c=0$ and $\mathbf{u}=0$. The system is quenched instantaneously to temperature $T$ and held at this temperature. For each value of $T$ we record the times at which 10\% martensite and 10\% pearlite form, as well as the time at which the austenite content drops below 10\%. 


\begin{figure}
\centering
\setlength{\unitlength}{1cm}
\begin{picture}(7.5,7)(.2,0)
\shortstack[c]{
\subfigure
{
    \label{128-0_0-TTT}
    \includegraphics[width=7.5cm]{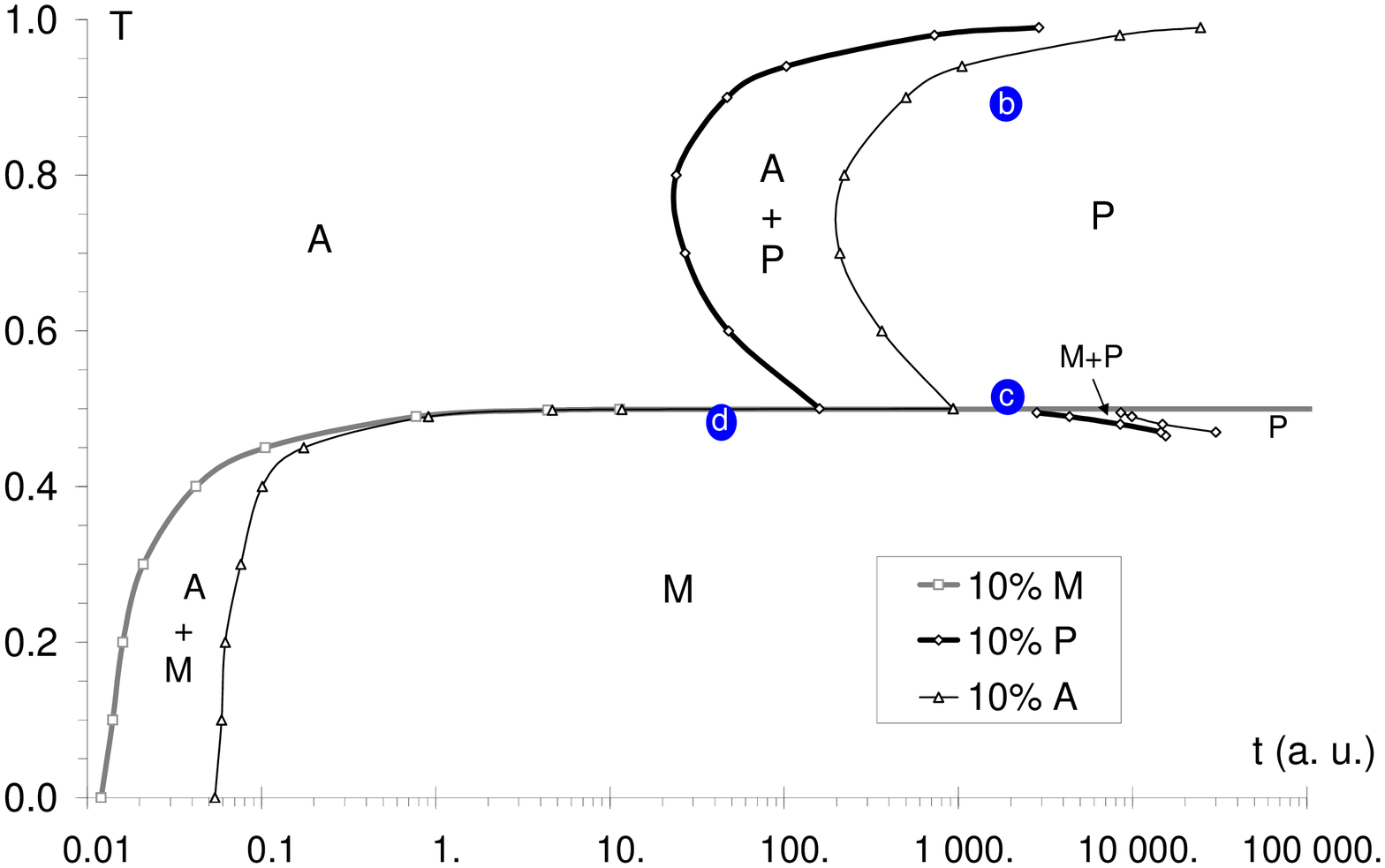}
\put(-0.8,4.1){\bf(a)}
}\vspace{-.2cm}\\
\subfigure
{
    \label{coarse-pearlite}
    \includegraphics[height=2.05cm]{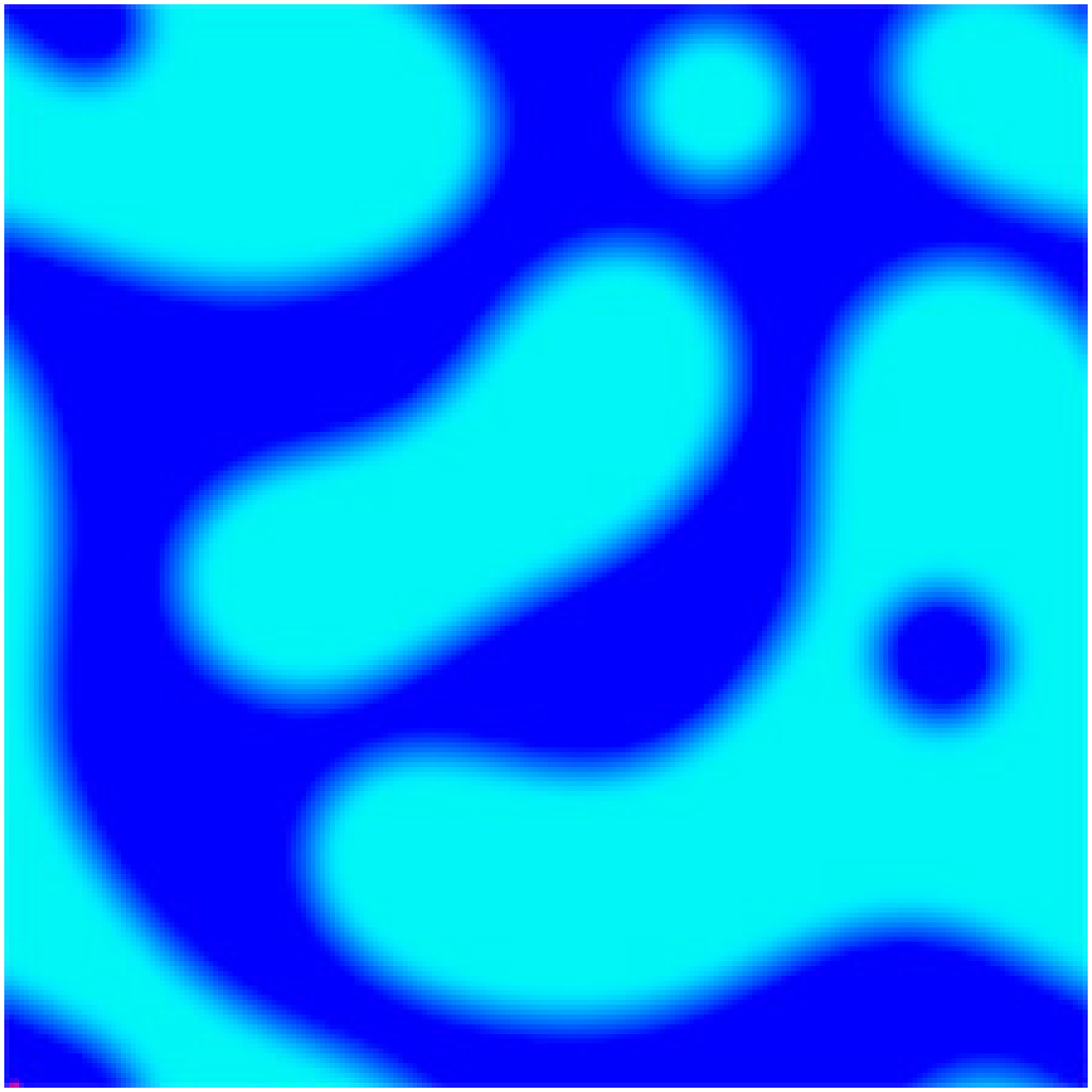}
\put(-0.65, 1.6){\color{white}\bf(b)}
}\quad
\subfigure
{
    \label{fine-pearlite}
    \includegraphics[height=2.05cm]{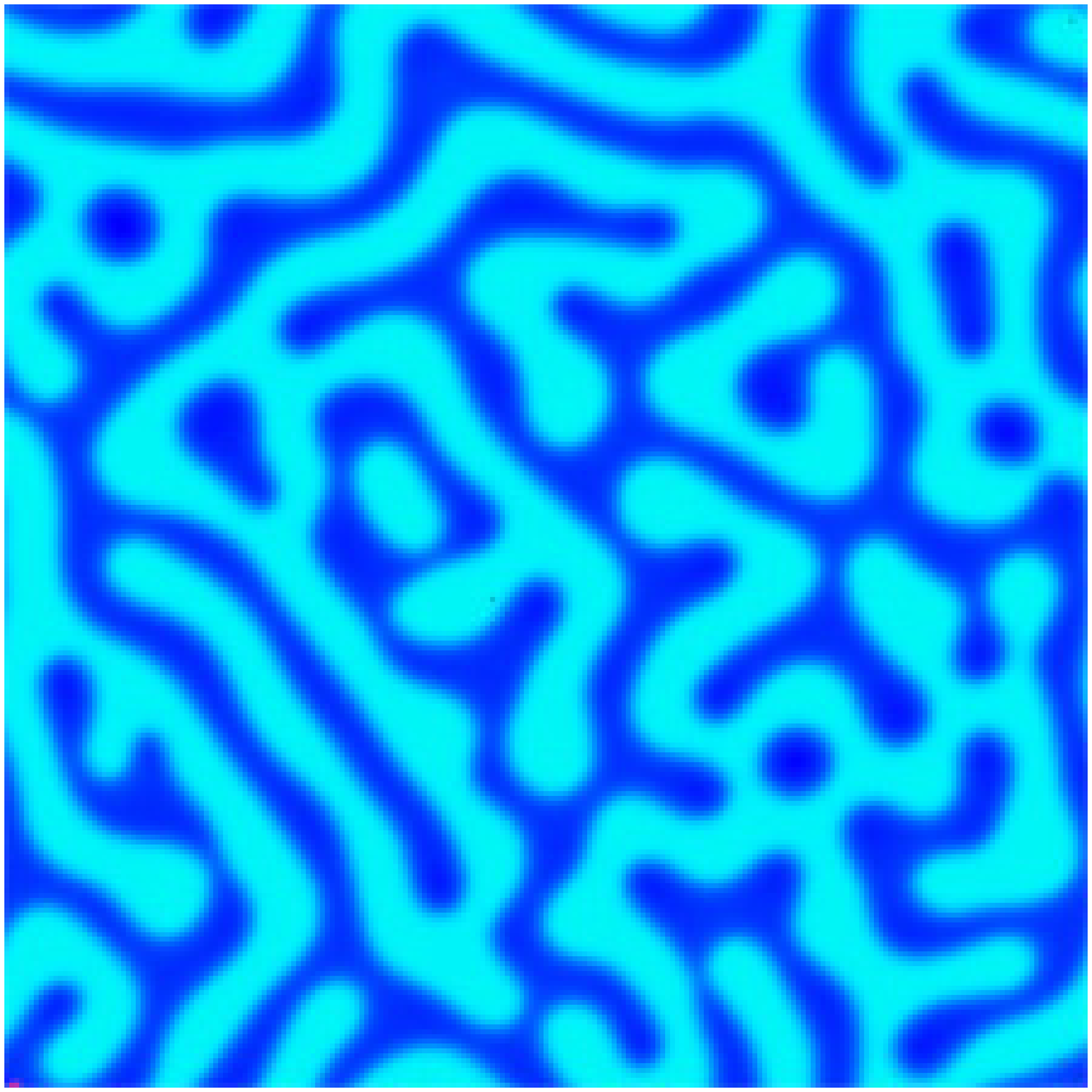}
\put(-0.65, 1.6){\color{white}\bf(c)}
}\quad
\subfigure
{
    \label{martensite}
    \includegraphics[height=2.05cm]{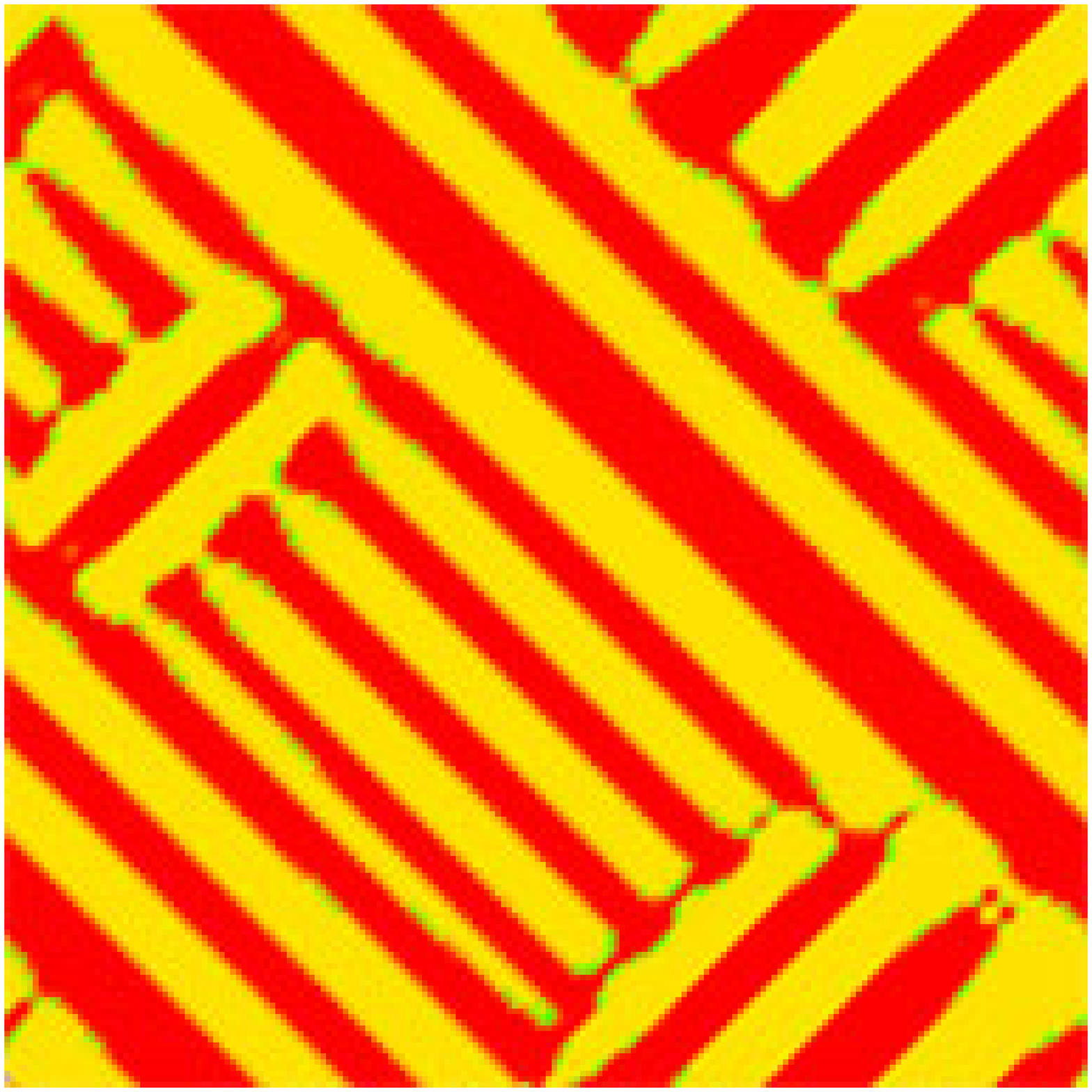}
\put(-0.65, 1.6){\bf(d)}
}
}
\end{picture}
	\caption{(color) (a) TTT diagram (A: austenite, M: martensite, and P: pearlite) and the microstructures at (b) $T=0.9$ and $t=2~000$, (c) $T=0.5$ and $t=2~000$, and (d) $T=0.49$ and $t=50$ for $x_{12} = x_{1c} = 0$. Red and yellow: martensite; light and dark blue: pearlite.}
\end{figure}

Figure~\ref{128-0_0-TTT} shows the resulting TTT 
diagram for the $x_{12} = x_{1c} = 0$ case. The austenite--pearlite phase transformation requires diffusion and therefore time. At low temperature, diffusion is slow and so is pearlite formation. Martensite on the other hand forms through a displacive mechanism, which does not require long-range motion of atoms. Figure~\ref{128-0_0-TTT} shows that at temperatures below about $T=0.5$ (the ``martensite start temperature'') austenite transforms to martensite (Fig.~\ref{martensite}) and pearlite forms only at higher temperature. At temperatures close to $T\p=1$ the undercooling is so small that pearlite formation is very slow. Pearlite formation is also delayed at temperatures close to $T=0.5$ because of the slow diffusion. Consequently there exists an intermediate temperature at which pearlite formation is the fastest and a C-curve can be observed around $T = 0.8$ in Fig.~\ref{128-0_0-TTT}. 
The typical length scale of pearlite depends on diffusion length and therefore on temperature: at low temperatures pearlite is fine \mbox{---Fig.~\ref{fine-pearlite}---} and it is coarser close to $T\p$ ---Fig.~\ref{coarse-pearlite}\mbox{---,} consistent with experimental observations. Note that the amounts of the two components of pearlite are equal because their compositions are symmetric with respect to that of austenite (Fig.~\ref{energy}).

\begin{figure}
\centering
\setlength{\unitlength}{1cm}
\begin{picture}(7.,2.05)(.2,0)
\subfigure
{
    \label{128-0_0-T0490(a)}
    \includegraphics[height=2.05cm]{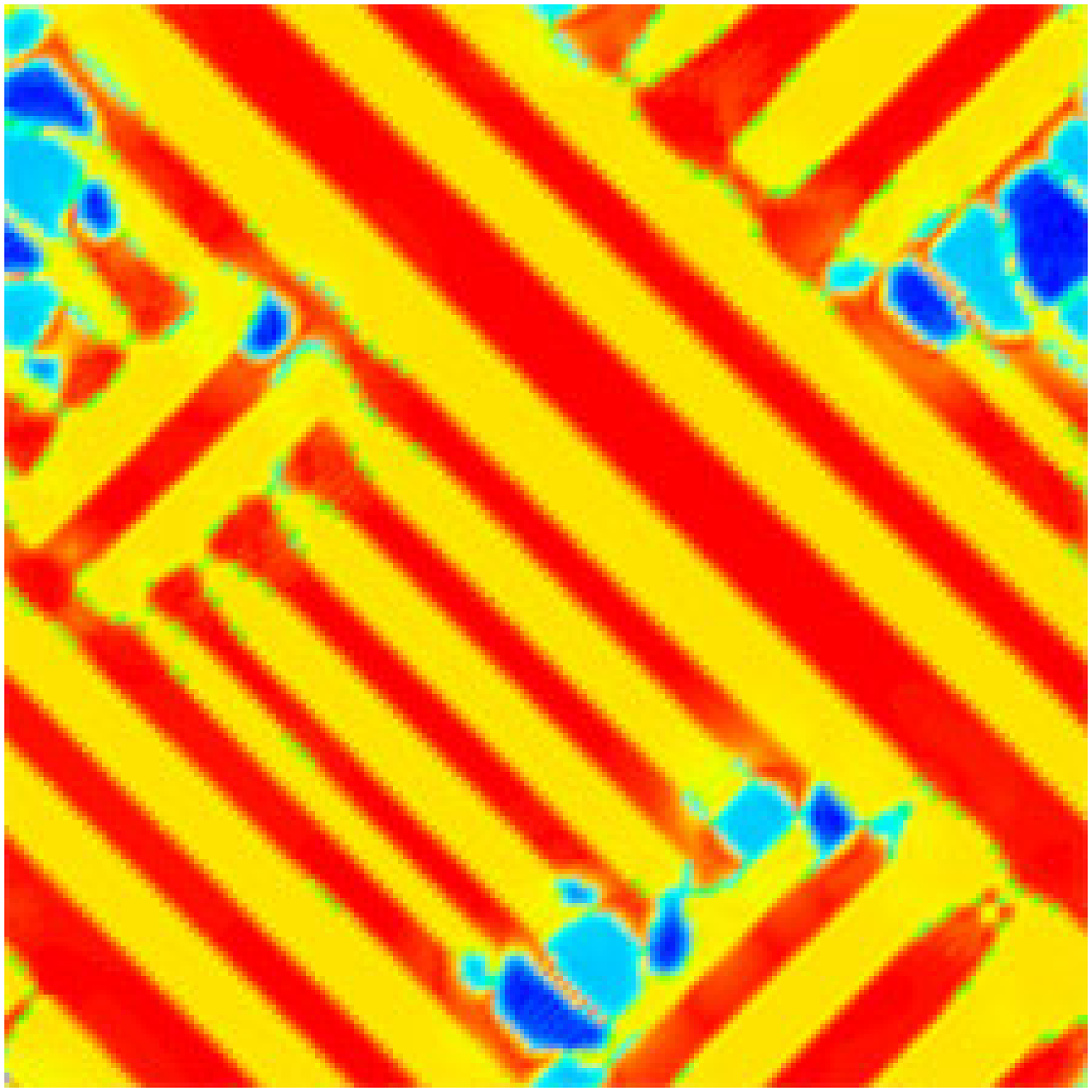}
\put(-0.65, 1.6){\bf(a)}
}\quad
\subfigure
{
    \label{128-0_0-T0490(b)}
    \includegraphics[height=2.05cm]{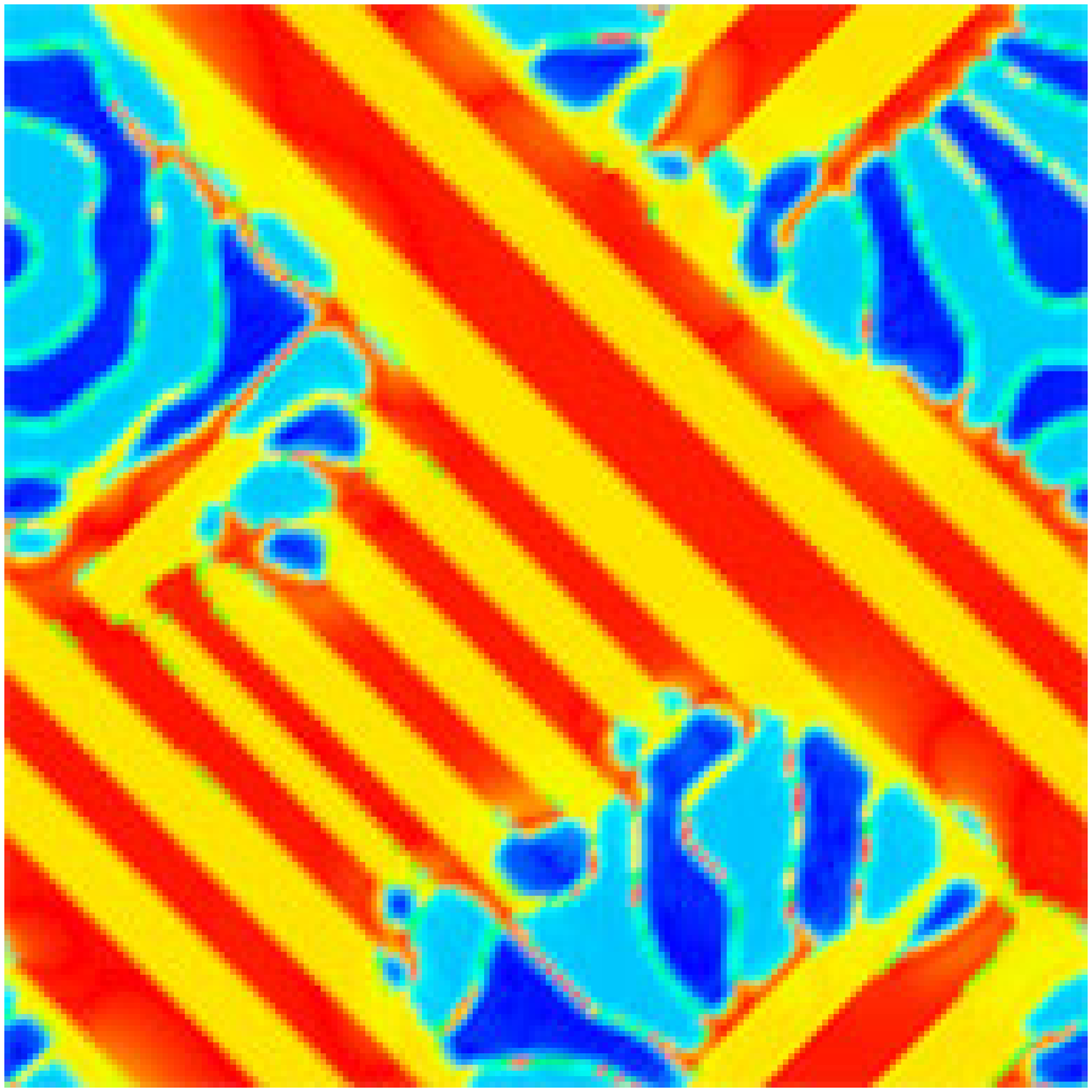}
\put(-0.6, 1.65){\color{white}\bf(b)}
}\quad
\subfigure
{
    \label{128-0_0-T0490(c)}
    \includegraphics[height=2.05cm]{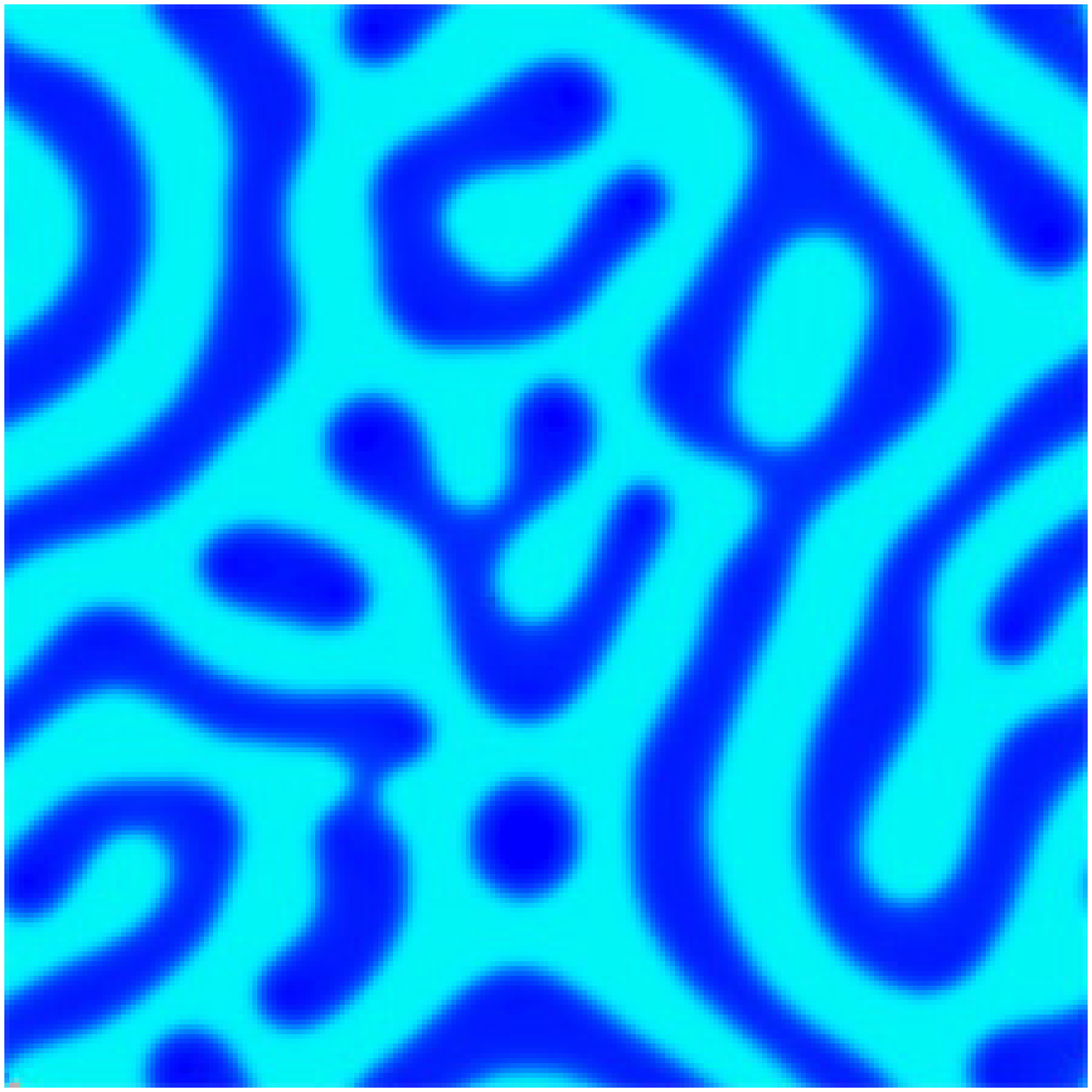}
\put(-0.65, 1.6){\color{white}\bf(c)}
}
\end{picture}
	\caption{\label{128-0_0-T0490}(color) Microstructures for $x_{12} = 0$ and $x_{1c} = 0$ at $T=0.49$ and (a) $t=4~000$, (b) $t=6~000$, and (c) $t=13~000$. Red and yellow: martensite; light and dark blue: pearlite.}
\end{figure}

Along with Fig.~\ref{martensite}, Fig.~\ref{128-0_0-T0490} shows the microstructure evolution at $T=0.49$ (just below the martensite start temperature). At $t=50$ only martensite is found, Fig.~\ref{martensite}. At $t=4~000$, pearlite has already started forming at the interfaces between the martensite variants aligned along $[1\,1]$ and those along $[1\,\bar{1}]$, Fig.~\ref{128-0_0-T0490(a)}, that is at interfaces which are relatively high in energy. For longer times pearlite grows at the expense of martensite (Fig.~\ref{128-0_0-T0490(b)}) and the latter finally disappears (Fig.~\ref{128-0_0-T0490(c)}). This appearance of pearlite in the martensite region occurs at progressively later times at lower temperatures.

As pearlite is the ground state and martensite is only metastable the latter forms only because pearlite formation is slow at this temperature. It is therefore expected that pearlite can eventually form in a martensitic system as this will decrease the energy. However there exists an energy barrier because, unlike austenite, 
martensite is metastable. This makes pearlite nucleation from martensite slower than nucleation from austenite and accounts for the discontinuity of the 10\% pearlite line across $T=0.5$ in Fig.~\ref{128-0_0-TTT}.
Note that the model cannot be expected to accurately estimate the time necessary for pearlite to form as the nucleation process is not properly described.

\begin{figure}
\centering
\setlength{\unitlength}{1cm}
\begin{picture}(8.5,4.9)(-.05,0)
\shortstack[c]{
\subfigure{
	\put(-0.2, 0.4){\rotatebox{90}{$x_{1c}=0$}}
    \label{cross-terms(a)}
    \includegraphics[height=2.cm]{128-0_0-T0490_t000050-sm}
	\put(-0.65, 1.55){\bf(a)}
	\put(-1.45,2.1){$t=50$}
}\subfigure{
	\put(-0.5,2.5){$x_{12}=0$}
    \label{cross-terms(b)}
    \includegraphics[height=2.cm]{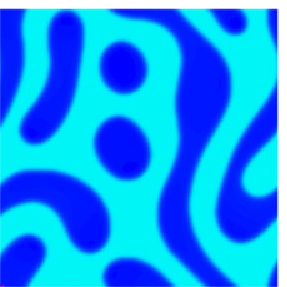}
	\put(-0.65, 1.55){\color{white}\bf(b)}
	\put(-1.7,2.1){$t=50~000$}
}
\subfigure{
    \label{cross-terms(c)}
    \includegraphics[height=2.cm]{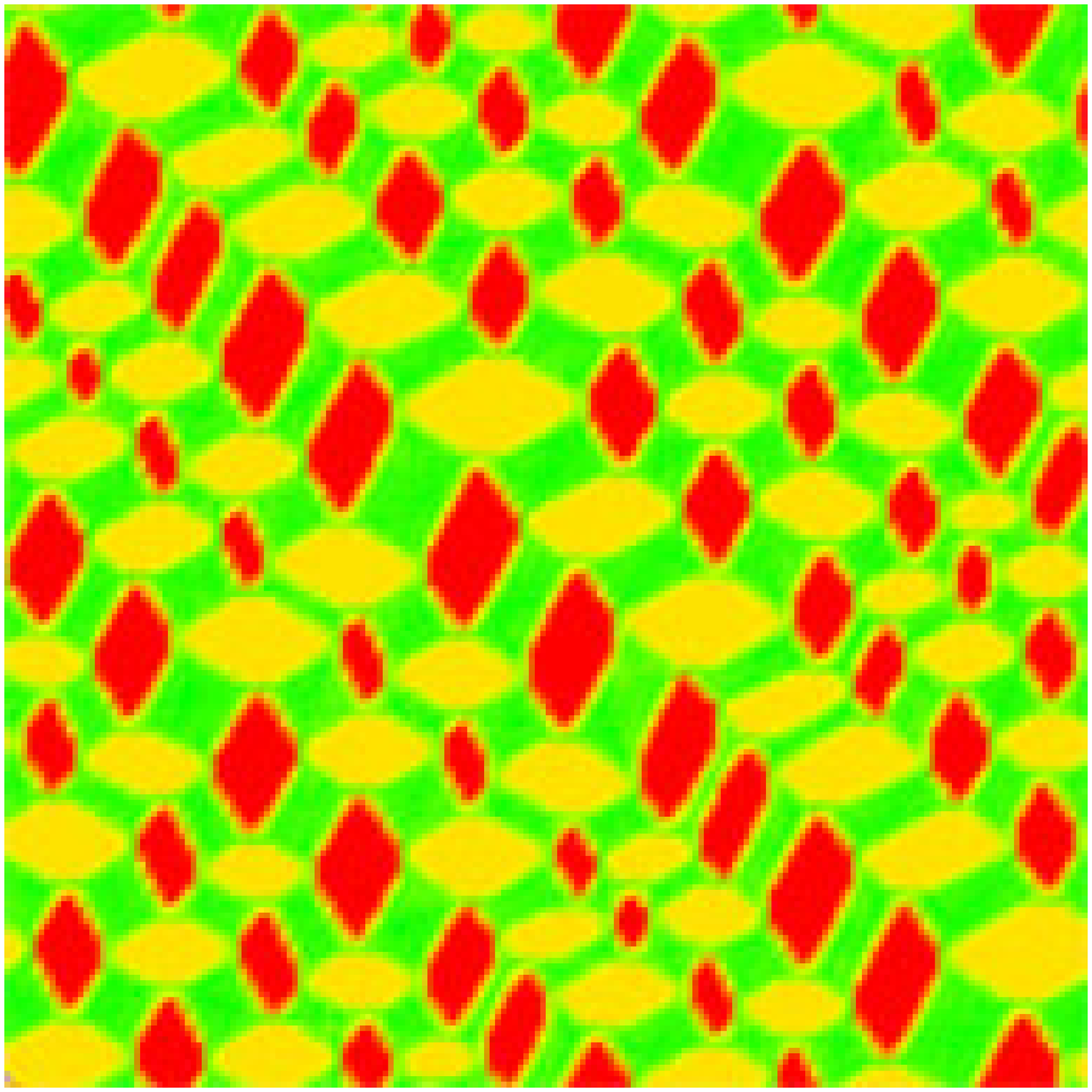}
	\put(-0.65, 1.55){\bf(c)}
	\put(-1.45,2.1){$t=50$}
}\subfigure{
	\put(-0.5,2.5){$x_{12}=1$}
    \label{cross-terms(d)}
    \includegraphics[height=2.cm]{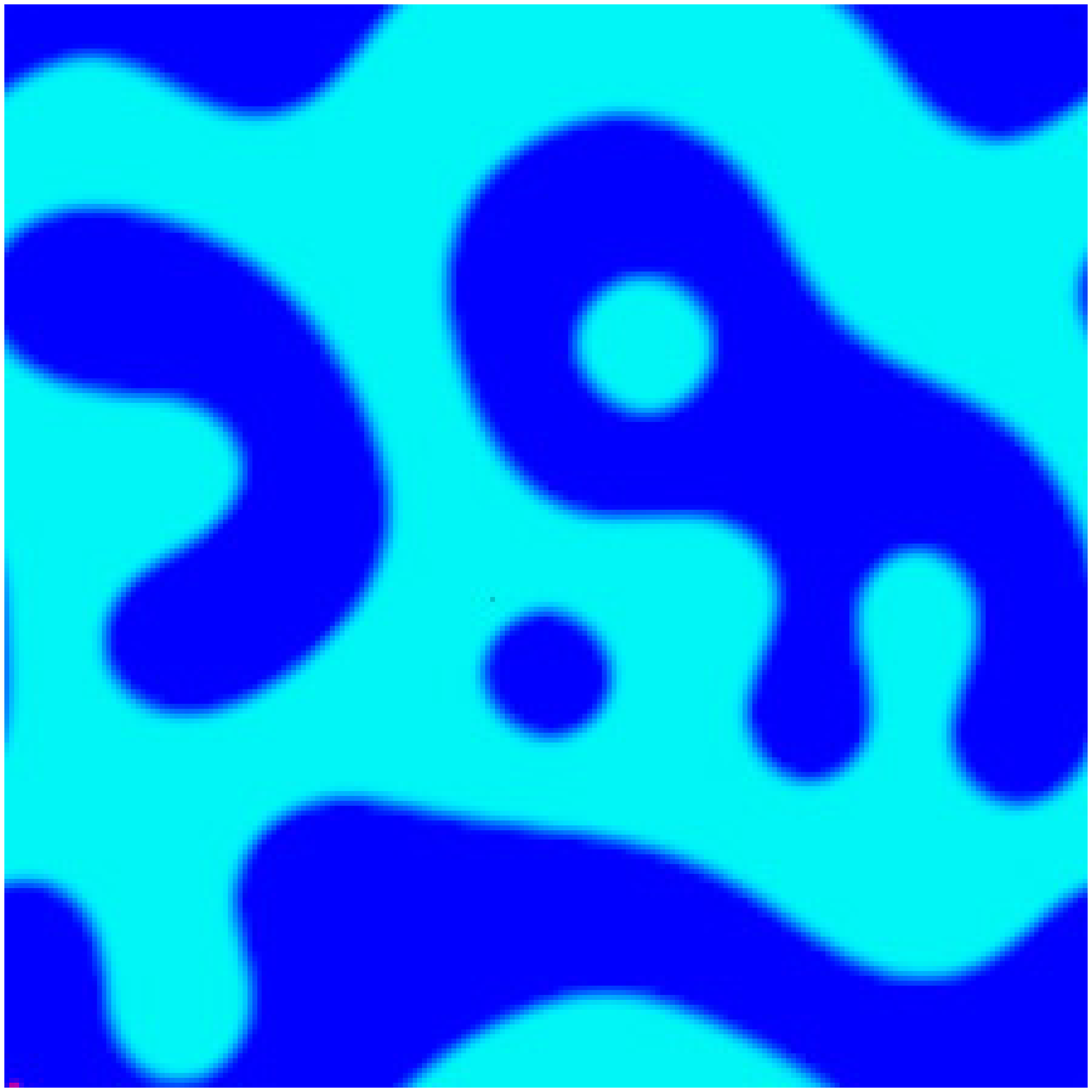}
	\put(-0.6, 1.55){\color{white}\bf(d)}
	\put(-1.7,2.1){$t=50~000$}
}\vspace{-.2cm}\\
\subfigure{
	\put(-0.2, 0.4){\rotatebox{90}{$x_{1c}=1$}}
    \label{cross-terms(e)}
    \includegraphics[height=2.cm]{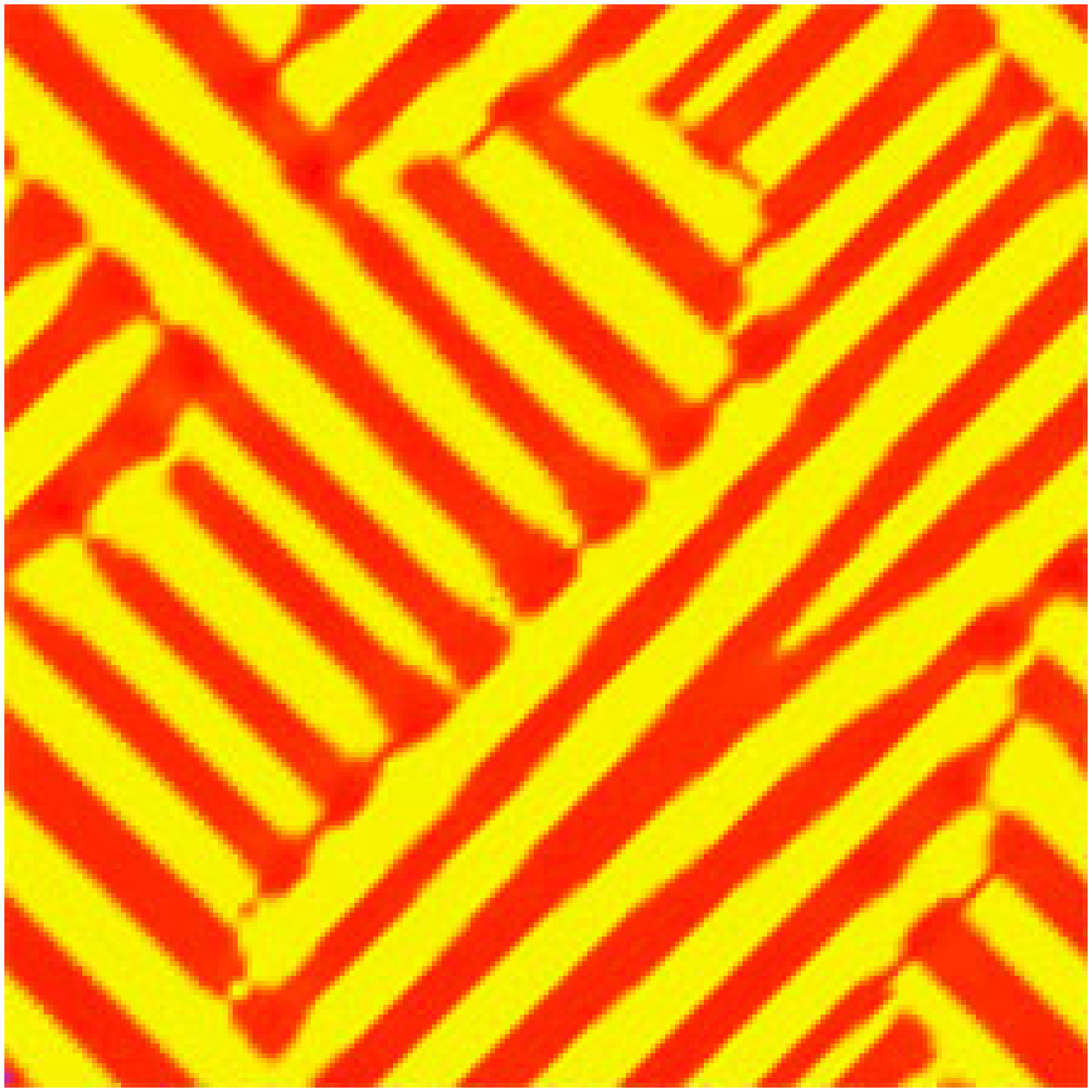}
	\put(-0.65, 1.55){\bf(e)}
}\subfigure{
    \label{cross-terms(f)}
    \includegraphics[height=2.cm]{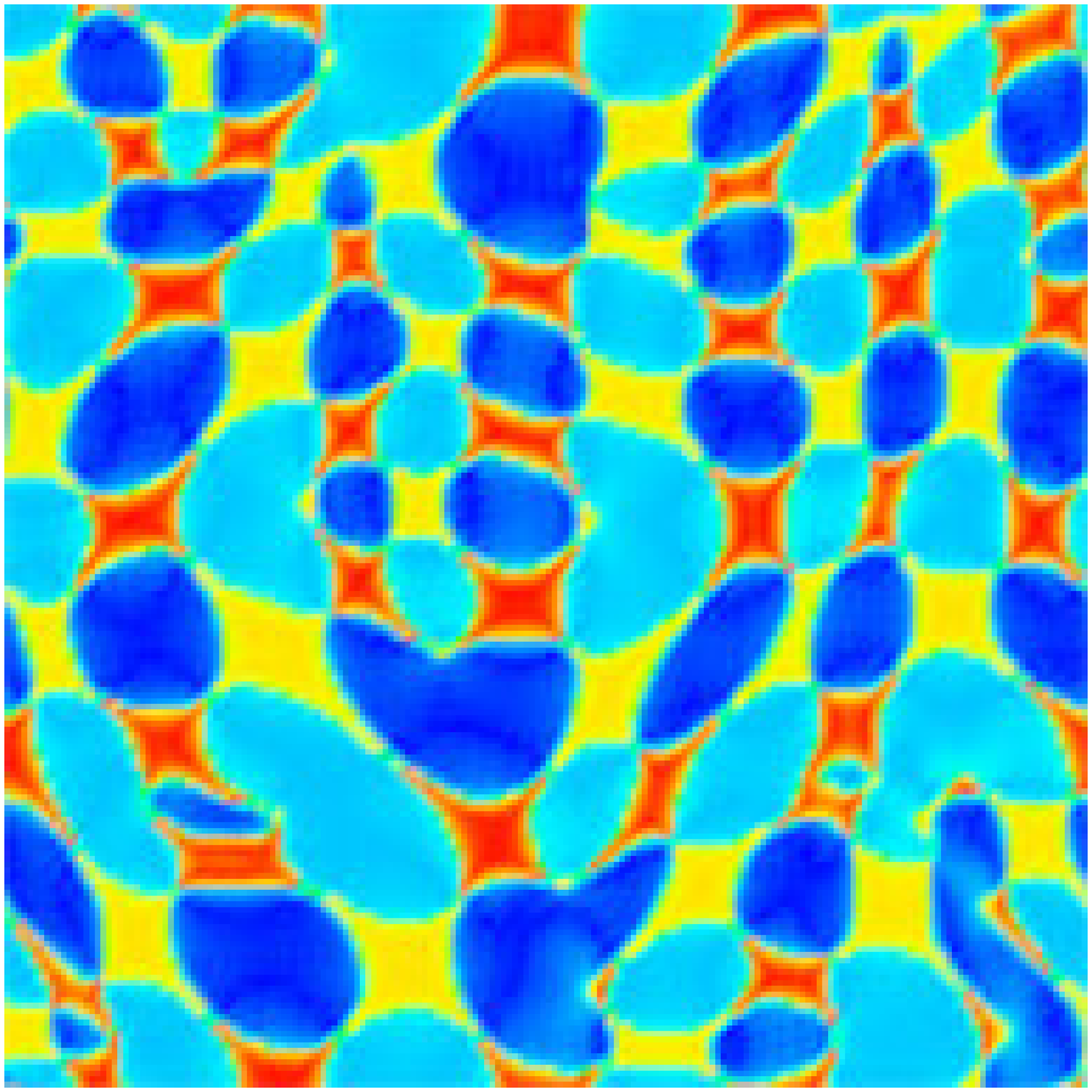}
	\put(-0.55, 1.55){\color{white}\bf(f)}
}\subfigure{
    \label{cross-terms(g)}
    \includegraphics[height=2.cm]{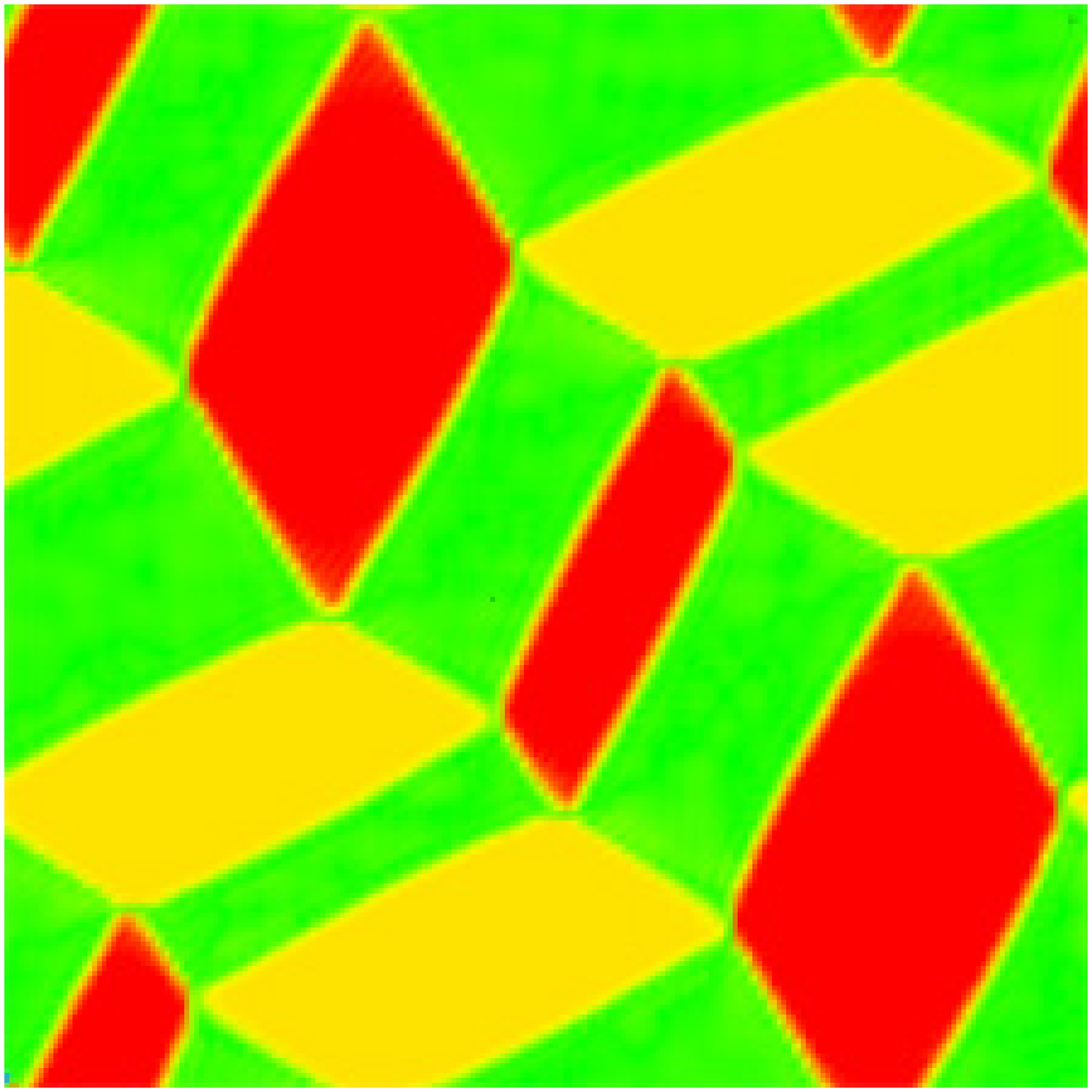}
	\put(-0.65, 1.55){\bf(g)}
}\subfigure{
    \label{cross-terms(h)}
    \includegraphics[height=2.cm]{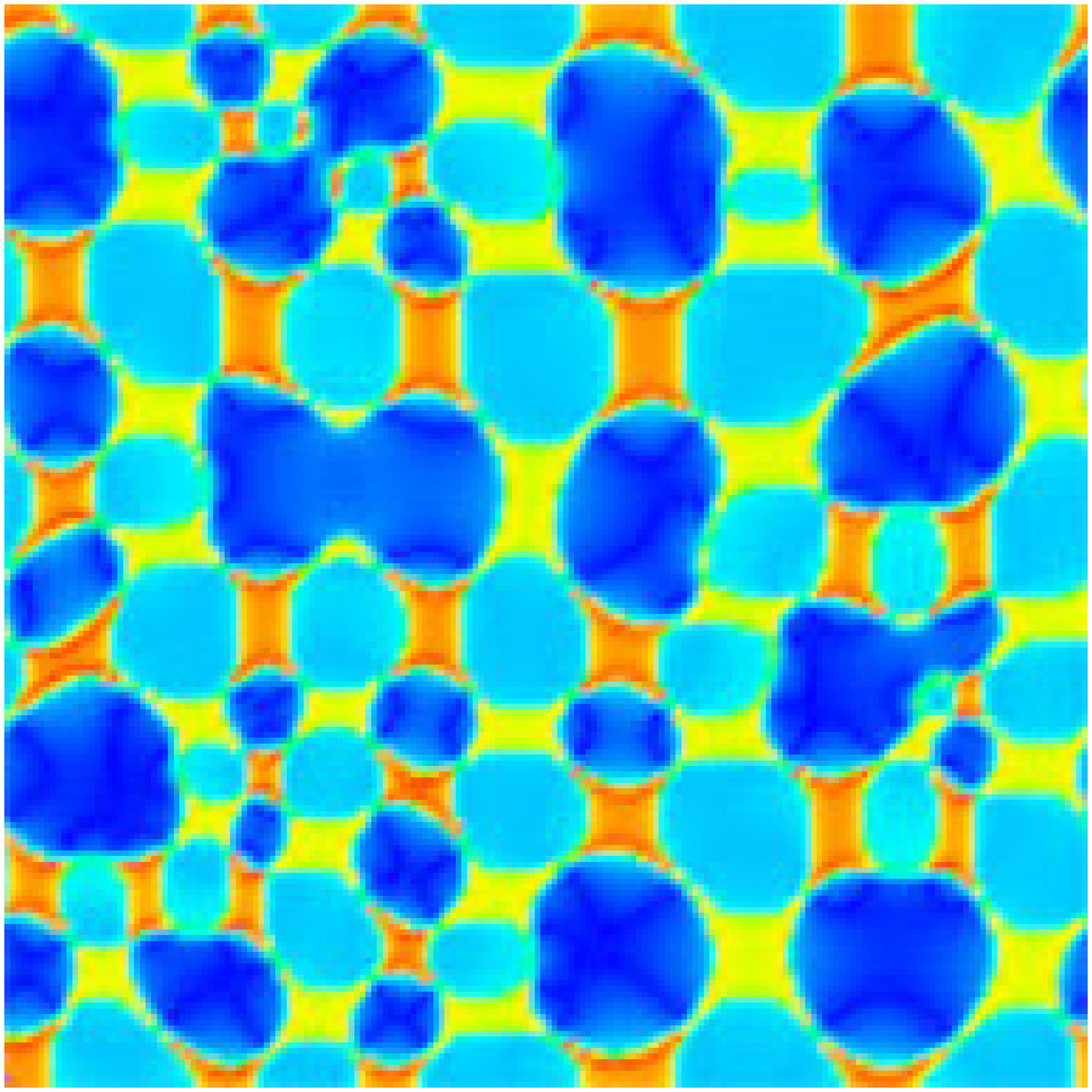}
	\put(-0.65,1.6){\color{white}\bf(h)}
}}
\end{picture}
	\caption{\label{cross-terms}(color) Microstructures at $T=0.49$. Top row: $x_{1c} = 0$, bottom row: $x_{1c} = 1$; two left columns: $x_{12} = 0$, two right columns: $x_{12} = 1$; columns 1 and 3: $t=50$, columns 2 and 4: $t=50~000$. Red and yellow: martensite; green: austenite; light and dark blue: pearlite.}
\end{figure}

So far we assumed that there was no bulk strain associated with pearlite or martensite formation, i.e.\ $x_{12}=x_{1c}=0$ in Eq.~(\ref{eq-g_el}). We now relax this constraint to study the effect of the term $x_{12}\,e_2^2$ of Eq.~(\ref{eq-g_el}) (keeping $x_{1c}$ set to zero). This term couples deviatoric strain (i.e.\ martensite) and hydrostatic strain: it associates a net volume change with the martensitic transformation. 
The effect of $x_{12}$ on the microstructure at $T=0.49$ can be seen in Fig.~\ref{cross-terms}. Whereas for low values of $x_{12}$ austenite completely transforms to martensite (e.g.\ Fig.~\ref{cross-terms(a)}), for larger values of $x_{12}$ martensite formation results in stresses so large that a system made of pure martensite would be unstable. Consequently, part of the system remains austenitic, Fig.~\ref{cross-terms(c)}. A similar microstructure was obtained by~\citet{Onuki-99}. This retained austenite transforms to pearlite, which then grows at the expense of martensite (Fig.~\ref{cross-terms(d)}).

The term $x_{1c}\,c$ in Eq.~(\ref{eq-g_el}), similar to Ref.~\onlinecite{Onuki-PRL-01}, corresponds to a lattice mismatch between the two components of pearlite (i.e.\ ferrite and cementite in the case of steel), such that the average lattice spacing of pearlite is that of austenite. Figure~\ref{cross-terms} shows the effect of $x_{12}$ and $x_{1c}$ on the microstructure. Since $x_{12}$ can stabilize austenite, at short times its value determines the microstructure: Figs.~\ref{cross-terms(a)} and~\ref{cross-terms(e)} (which correspond to $x_{12}=0$) show pure martensite whereas in Figs.~\ref{cross-terms(c)} and~\ref{cross-terms(g)} ($x_{12}=1$) there is a mixture of martensite and austenite. The long-time microstructures on the other hand are controlled by the value of $x_{1c}$: in Figs.~\ref{cross-terms(b)} and~\ref{cross-terms(d)} (corresponding to $x_{1c}=0$) there is only pearlite whereas Figs.~\ref{cross-terms(f)} and~\ref{cross-terms(h)} ($x_{1c}=1$) show a mixed pearlite--martensite microstructure (which does not disappear in longer simulations). $x_{1c}$ stabilizes martensite at long times: when $x_{1c}=1$ a purely pearlitic system would not be stable because of the large stresses it would generate. This is a kind of stress-induced martensitic transformation.

As expected~\cite{Sapriel75} martensite--martensite interfaces are oriented along $\langle 1\,1 \rangle$. 
The orientations of martensite--austenite and martensite--pearlite interfaces are different. 
For an interface at an angle $\theta$ with respect to $[1\,0]$, elastic compatibility requires that $\cos 2\theta = \delta e_1/\delta e_2$ where $\delta e_1$ and $\delta e_2$ are the variations of $e_1$ and $e_2$ across the interface. The values of $e_1$ and $e_2$ in Fig.~\ref{cross-terms(g)} give $\theta \approx \pm 30^\text{o}$ or $\theta \approx \pm 60^\text{o}$, which is consistent with the interface orientations observed in the figure. In Fig.~\ref{cross-terms(h)} the rounded shape of pearlite grains aims at minimizing interface energy; therefore the orientation of the interfaces is not determined by elasticity alone.

\begin{figure}
\centering
\setlength{\unitlength}{1cm}
\begin{picture}(8.3,7.6)(.2,0)
\shortstack[c]{
\subfigure
{
	\label{128-1_1-TTT}
	\includegraphics[width=7.5cm]{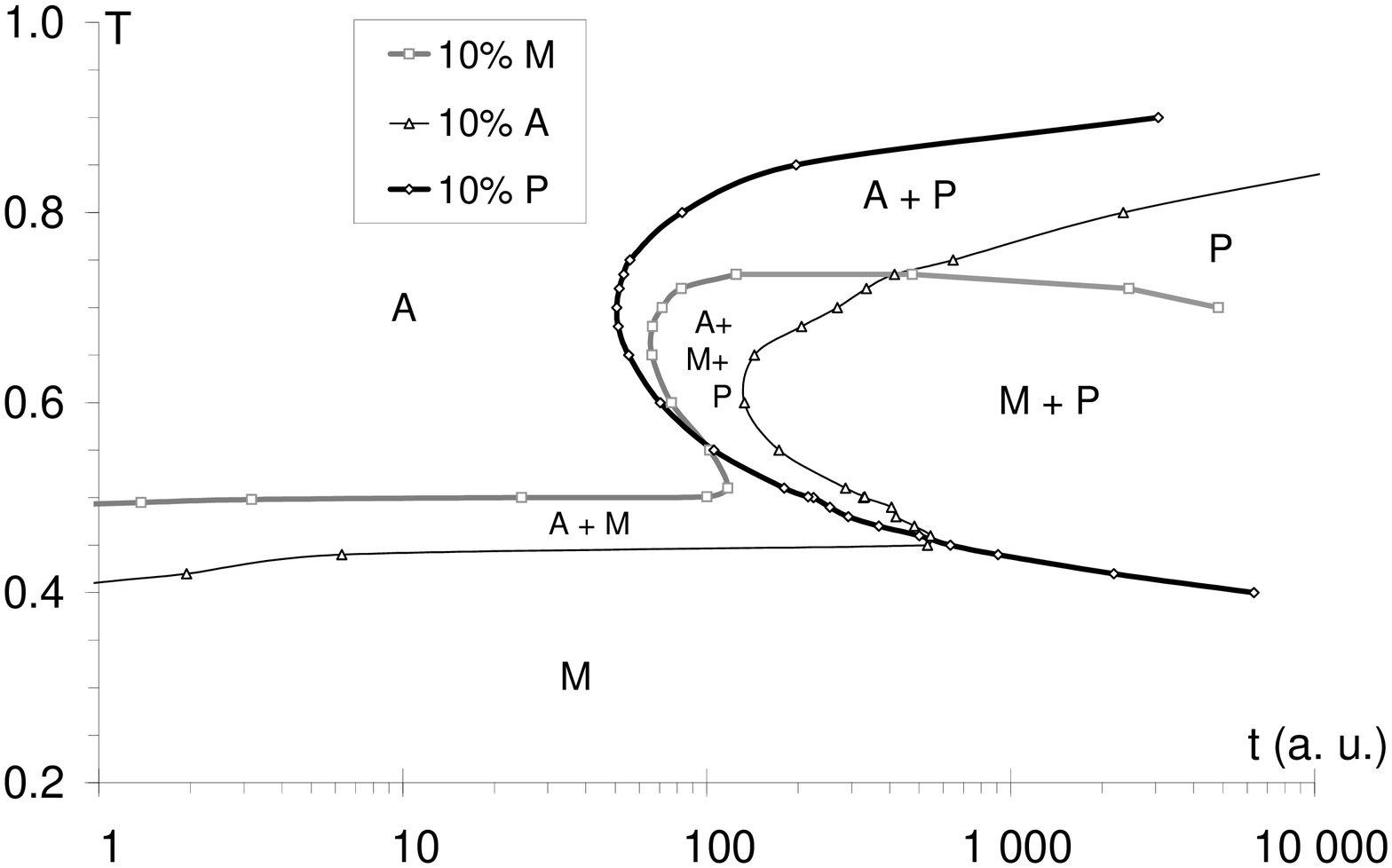}
\put(-0.8,4.1){\bf(a)}
}\vspace{-.2cm}\\
\subfigure
{
    \label{128-1_1-T0490_t001000}
    \includegraphics[height=2.65cm]{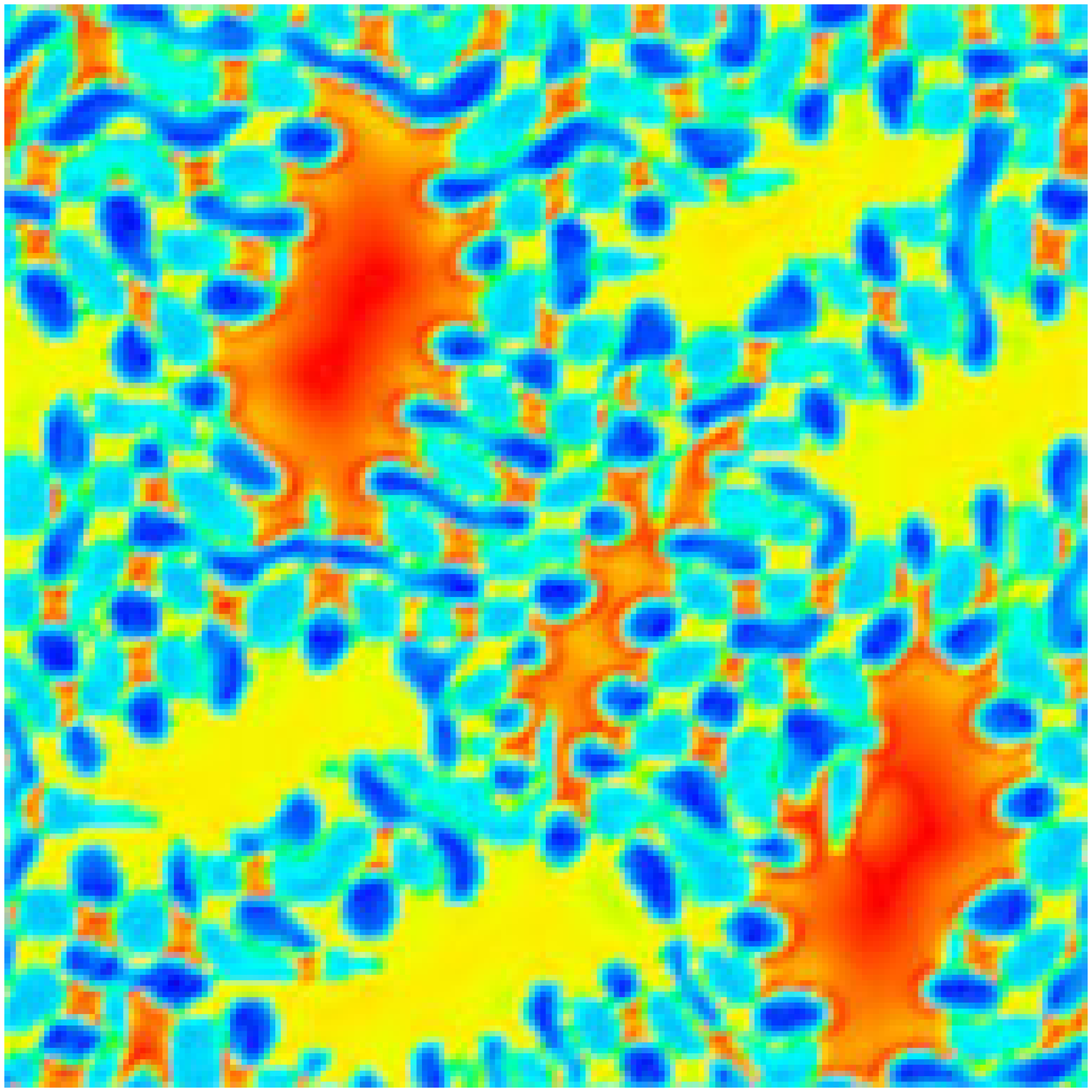}
\put(-0.65,2.2){\bf(b)}
}
\subfigure
{
    \label{128-1_1-T0490_t001500}
    \includegraphics[height=2.65cm]{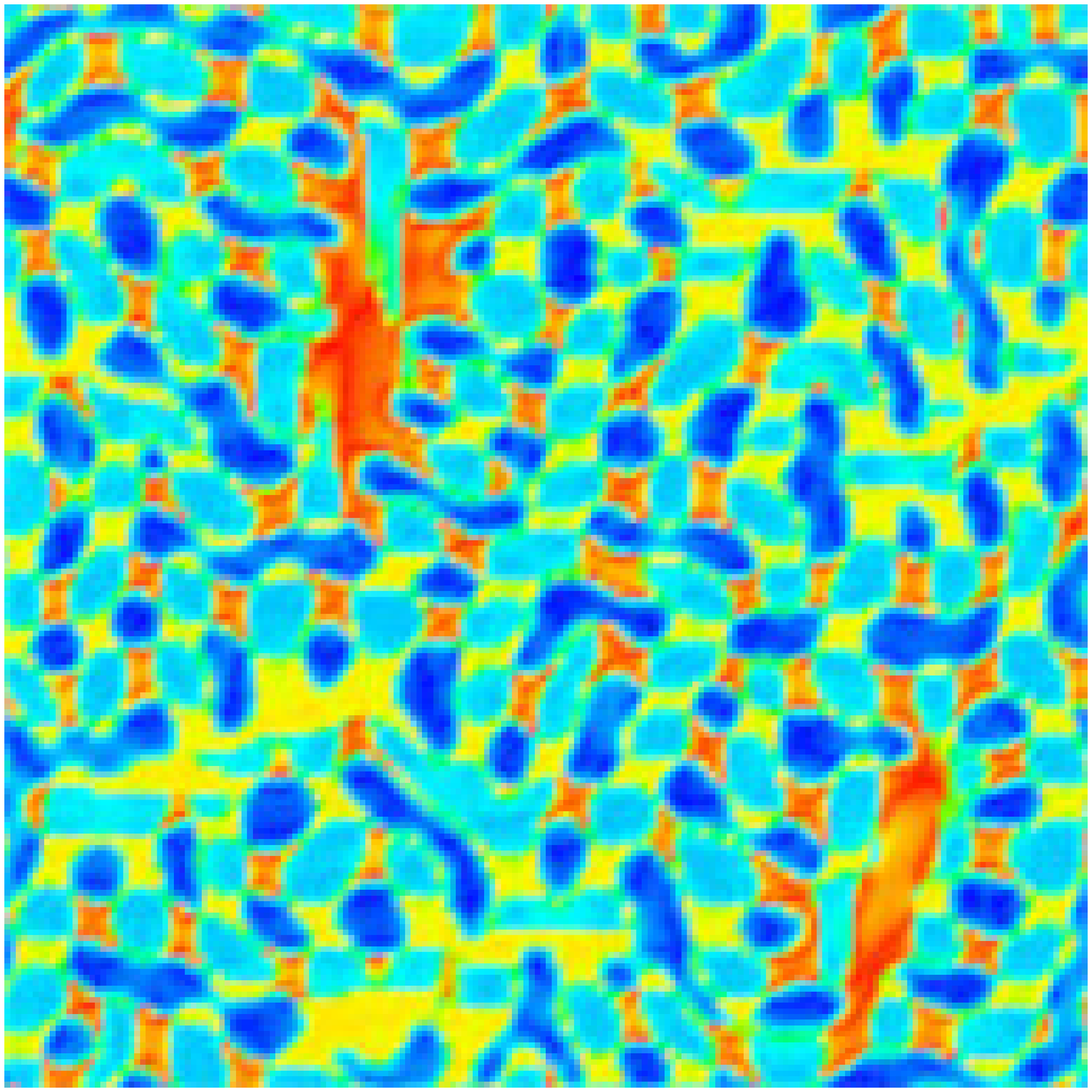}
\put(-0.65,2.2){\bf(c)}
}
\subfigure
{
    \label{128-1_1-T0490_t002000}
    \includegraphics[height=2.65cm]{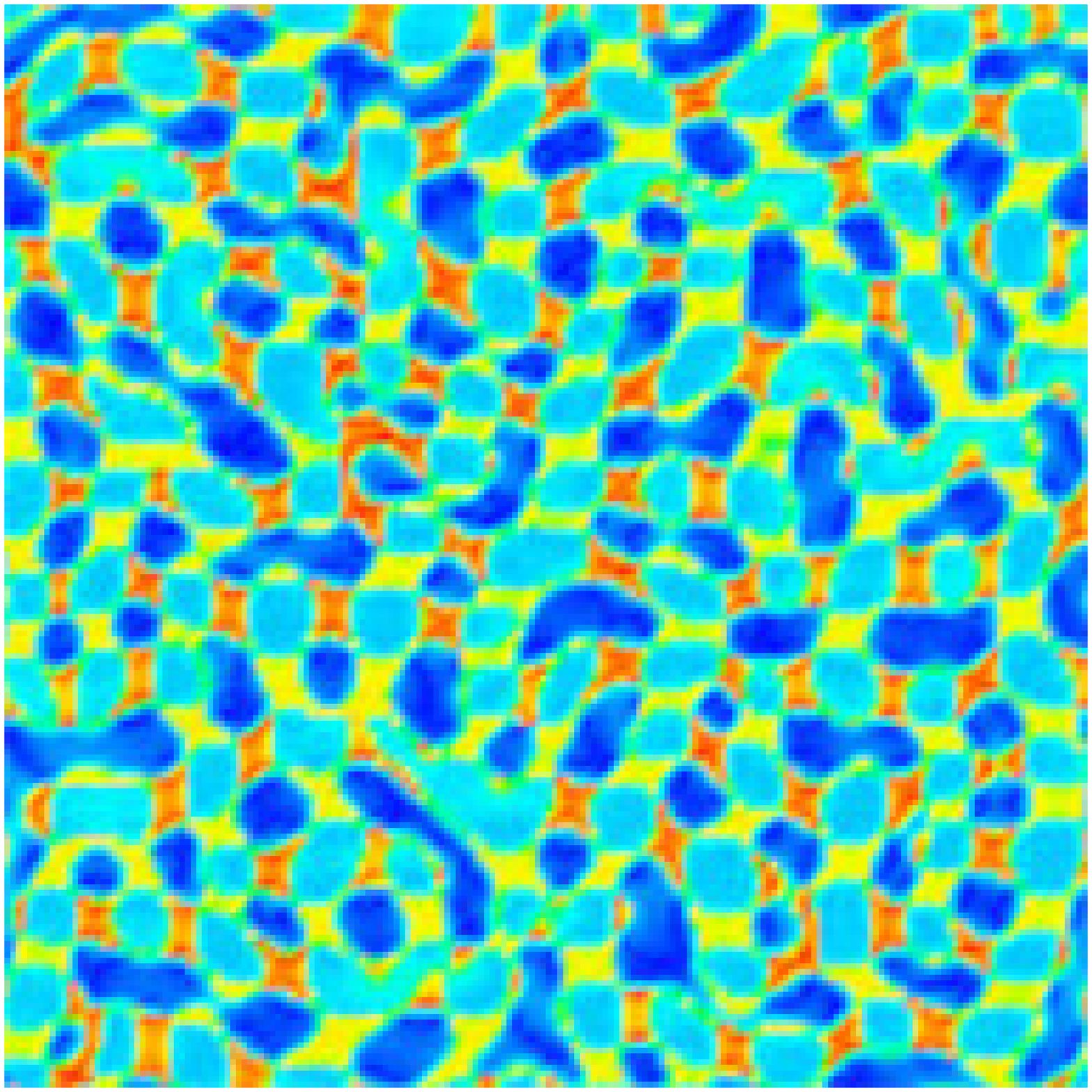}
\put(-0.65,2.2){\bf(d)}
}}
\end{picture}
	\caption{\label{128-1_1}(color) (a): TTT diagram for $x_{12}=x_{1c}=1$ (A: austenite, M: martensite, and P: pearlite). Microstructures at $T=0.49$: (b) $t=1~000$, (c) $t=1~500$, and (d) $t=2~000$ (red and yellow: martensite; light and dark blue: pearlite).}
\end{figure}

Figure~\ref{128-1_1} focuses on $x_{12}=x_{1c}=1$.  Several features of the TTT diagram shown in Fig.~\ref{128-1_1-TTT} are different from the TTT diagram obtained for $x_{12}=x_{1c}=0$, Fig.~\ref{128-0_0-TTT}. There exists a martensite finish temperature (around $T=0.45$) due to retained austenite. Martensite can be found above $T=0.5$ because ---due to underlying dilation strains and elastic compatibility--- a mixture of pearlite and martensite is more stable than pure pearlite (this remains true up to $T \approx 0.73$). This splitting of the pearlite region in the TTT diagram is akin to the pearlite--bainite transition in steel (however, the microstructure shown in Fig.~\ref{cross-terms(h)} is different from that of bainite). When $x_{12}=x_{1c}=0$, below $T=0.5$ pearlite can nucleate heterogeneously at the interface between martensite variants; this process is rather slow. When $x_{12}=x_{1c}=1$, even below $T=0.5$ some austenite remains and pearlite formation proceeds in this retained austenite. Since austenite is unstable there is no nucleation barrier and there is no discontinuity of the 10\% pearlite line (the pearlite nucleation is slower at lower $T$ only because the diffusivity is lower). 

Figures~\ref{cross-terms(g)}, \ref{128-1_1-T0490_t001000}--\ref{128-1_1-T0490_t002000}, and~\ref{cross-terms(h)} show the time evolution of the microstructure at $T=0.49$ for $x_{12}=x_{1c}=1$. By $t = 1~000$, pearlite nucleated in the retained austenite and new martensite formed where there was none before (pearlite cannot form alone because of the stresses it generates), Fig.~\ref{128-1_1-T0490_t001000}. After austenite disappears pearlite grows at the expense of the large grains of ``primary'' martensite and makes them ``rotate'' towards $\langle 0\,1 \rangle$, Fig.~\ref{128-1_1-T0490_t001500}. From large martensite grains alternating with areas completely devoid of martensite at $t=50$, Fig.~\ref{cross-terms(g)}, the system evolves to a state where martensite is more homogeneously distributed, Fig.~\ref{128-1_1-T0490_t002000} at $t=2~000$, through a double mechanism of martensite formation and martensite destruction, and finally to a coarsened microstructure, Fig.~\ref{cross-terms(h)}. The feature size thus goes from large to small to medium with increasing time. Consequently the ``final'' structure is independent of the initial one, both in terms of grain size and interface orientation (compare Fig.~\ref{cross-terms(h)} to Fig.~\ref{cross-terms(g)}).

A new Ginzburg--Landau approach has been proposed to study alloys ---such as steels--- which can undergo displacive as well as diffusive transformations.
It captures the important features of TTT diagrams and microstructures 
and sheds some light on the role of the interplay between the two types of transformations in stabilizing mixed microstructures.
The existence of a martensite finish temperature (i.e.\ of retained austenite) is due to an hydrostatic strain associated with the martensitic transformation.
When a strain is associated with pearlite formation, martensite and pearlite coexist at intermediate temperatures, i.e.\ in the region of the TTT diagram where bainite is typically found in steels. 
The model also shows that in these mixed microstructures the habit planes can be different from the pure martensite case.

\acknowledgments
We thank David Srolovitz and Srikanth Vedantam for useful discussions.

\bibliography{mart}
\bibliographystyle{apsrev}

\end{document}